\documentclass[11pt,final]{iopart}

\newcommand{\U}{\hat U}
\newcommand{\Ug}{{\hat U_\text{geo}}}
\newcommand{\Us}{{\hat U_\text{sp}}}
\newcommand{\Uf}{{{\hat U}_\text{sp,tot}}}
\newcommand{\TR}{{\hat {\mathcal{T}}}}
\newcommand{\be}{\begin{equation}}
\newcommand{\ee}{\end{equation}}
\newcommand{\bp}{\mathbf{p}}
\newcommand{\bq}{\mathbf{q}}
\newcommand{\bn}{\mathbf{n}}
\newcommand{\bS}{\mathbf{S}}
\newcommand{\bL}{\mathbf{L}}
\newcommand{\bJ}{\mathbf{J}}
\newcommand{\bC}{\mathbf{C}}
\newcommand{\bA}{\mathbf{A}}
\newcommand{\bB}{\mathbf{B}}
\newcommand{\by}{\mathbf{y}}
\newcommand{\bsigma}{\boldsymbol{\sigma}}
\newcommand{\Id}{\mathbf{1}}
\newcommand{\R}{\mathcal{R}}
\newcommand{\I}{\mathcal{I}}
\newcommand{\HS}{\mathcal{H}}
\newcommand{\ket}[1]{| #1\rangle}
\newcommand{\bra}[1]{\langle #1 |}
\newcommand{\FS}{\mathbb{F}}
\newcommand{\K}{\mathcal{K}}
\newcommand{\bpsi}{\boldsymbol{\psi}}

\renewcommand{\paragraph}[1]{\subsubsection{#1}}

\expandafter\let\csname equation*\endcsname\relax
\expandafter\let\csname endequation*\endcsname\relax
\usepackage{amsmath,amssymb}

\let\Re\relax
\DeclareMathOperator{\Re}{Re}
\let\Im\relax
\DeclareMathOperator{\Im}{Im}

\usepackage{graphicx}
\usepackage{xcolor}
\usepackage{cite}

\begin{document}

\title{Trace formula for quantum chaotic systems with geometrical symmetries and spin}
\author{Vaios Blatzios$^1$, Christopher H. Joyner$^2$, Sebastian M\"{u}ller$^1$ and Martin Sieber$^1$}
\address{
$^1$School of Mathematics, University of Bristol, Fry Building, Woodland Road, Bristol BS8 1UG, United Kingdom \\
$^2$School of Mathematical Sciences, Queen Mary University of London, London, E1 4NS, United Kingdom}


\begin{abstract}
We derive a Gutzwiller-type trace formula for quantum chaotic systems that accounts for both particle spin precession and discrete geometrical symmetries. This formula generalises previous results that were obtained either for systems with spin  \cite{Bolte-1998, Bolte-1999a} or for systems with symmetries \cite{Robbins-1989, Lauritzen-1991}, but not for a combination of both.
The derivation requires not only a combination of methodologies for these two settings, but also the treatment of new effects
in the form of double groups and spin components of symmetry operations. The resulting trace formula expresses the level density of subspectra
associated to irreducible representations of the group of unitary symmetries in terms of periodic orbits in the system's fundamental domain. 
We also derive a corresponding expression for the spectral determinant. In a follow-up paper \cite{Blatzios-2024b} we will show that our formula
allows to study the impact of geometrical symmetries and spin on spectral statistics.

\end{abstract}

\section{Introduction}

Semiclassical methods allow to approximate properties of quantum systems by relating them to trajectories of the corresponding classical system. A key example is the propagator of a quantum system that can be approximated by a sum over classical trajectories \cite{VanVleck-1928, Gutzwiller-1970, Gutzwiller-1971}. Analogous approximations exist for its Laplace transform, the Green function. 
Here, we are interested in trace formulas that express the level density of a quantum chaotic system in terms of their classical periodic orbits. For general chaotic systems such a trace formula was derived by Gutzwiller \cite{Gutzwiller-1970,Gutzwiller-1971} (see also Balian and Bloch's work \cite{Balian-1970, Balian-1972} for chaotic billiards). Selberg's earlier derivation of a trace formula for compact Riemannian surfaces \cite{Selberg-1956} formed the basis of a large body of rigorous work. 

Gutzwiller's trace formula can be used to explain why the spectral statistics of fully chaotic quantum systems, studied in the semiclassical limit, take a universal form \cite{Hannay-1984,Berry-1985,Argaman-1993,Sieber-2001,Muller-2005,Muller-2009}. As conjectured in \cite{Bohigas-1984}, these statistics align with predictions
from random-matrix theory (RMT), and in the absence of symmetries their form depends only on the behaviour of the system under time reversal. In systems without time-reversal
invariance the relevant random matrix ensemble is the Gaussian Unitary Ensemble (GUE). For time-reversal invariant systems one has to use the Gaussian Orthogonal
Ensemble (GOE) or the Gaussian Symplectic Ensemble (GSE) instead, depending on whether the time-reversal operator squares to $1$ or $-1$, respectively. Similar methods can be applied to quantum chaotic transport \cite{Richter-2000,Richter-2002,Muller-2007a}. 

In the present article, we derive a trace formula for chaotic systems that have both geometrical symmetries and spin. This work is motivated by studies 
of the effect of symmetries on spectral statistics. If a system has geometrical symmetries the spectrum falls into parts associated to irreducible representations
(irreps) of the symmetry group.  The statistics inside these subspectra depend not only on the time-reversal properties of the system but also on the type of irrep 
\cite{Wigner-1959,Keating-1997,Joyner-2012,Joyner-2013}.
For example, Leyvraz, Schmit and Seligman \cite{Leyvraz-1997} observed statistics according to the GUE for a subspectrum of a system with three-fold
rotation symmetry, even though it is invariant under a time-reversal operator squaring to 1.

As in the case of quantum chaotic systems without geometrical symmetries, one can use semiclassical methods to explain the type of spectral statistics that one finds
in each subspectrum. This was done in \cite{Joyner-2012, Joyner-2013} which extended the work \cite{Keating-1997} that used the so-called diagonal approximation. 
However, these semiclassical approaches have not been applied yet to systems with symmetries where the time-reversal operator squares to $-\Id$, which typically 
requires the incorporation of spin. It is the aim of our work to fill this gap. In the present paper we focus on the derivation of the trace formula for cases with
symmetries and spin, setting the stage for a study of spectral statistics in a follow-up paper \cite{Blatzios-2024b}.

To obtain a trace formula for systems with discrete symmetries and spin, we extend the work of Robbins \cite{Robbins-1989} who considered symmetries 
(see \cite{Lauritzen-1991} for an alternative approach), and Bolte and Keppeler \cite{Bolte-1998, Bolte-1999a} who considered spin.
In Robbins' case the notion of a fundamental domain is crucial. If the geometrical symmetries form a finite group of order $n$, the position space can be divided
into $n$ equivalent parts that are mapped into each other by application of the group elements. These equivalent parts define the fundamental domain of the
system, and the trace formula in \cite{Robbins-1989} is formulated for a restriction of the system to this domain with appropriate boundary conditions
(see Section \ref{subgreen}).

Combining the effects of geometrical symmetries with the effects of spin into a single trace formula requires the consideration of two important new aspects.
The first, double groups, arises because for systems with half-integer spin, rotation by $2\pi$ has to be considered as an element of the symmetry group
that acts trivially in position space but not in spin space (see Section \ref{sec:rotations}). 
The second new aspect is that, when the orbits are folded back into the fundamental domain to account for the presence of symmetries, this also affects the spin precession.

The result is the formula (\ref{trace}) below, which expresses the level density $\rho_{\alpha}(E)$ of an irrep $\alpha$ in
terms of periodic orbits $p$ in the fundamental domain. This applies to systems of both integer and half-integer spin, and we describe how the formula, specifically
the spin component $\tr(d_p)$ is adapted in each case in Section \ref{Sec: Symmetry reduced trace}.
We have that\begin{equation}\label{trace}
\rho_{\alpha}(E) \approx \bar{\rho}_{\alpha}(E) + \frac{1}{\pi \hbar}\Re \sum_{p}\chi_{\alpha}(g_p)\tr(d_p) A_p e^{iS_p(E)/\hbar},
\end{equation}
where $\bar\rho_\alpha(E)$ is the average level density (discussed further in Subsection \ref{Weyl}). The contribution of each orbit depends on its classical action $S_p=\int{\bp}\cdot d \bq$ (with $\bq$ and $\bp$ denoting position and momentum) and stability amplitude $A_p$, which can be written as $A_p = \frac{T_p^\text{prim} e^{-i\mu_p \pi/2}}{\sqrt{\det(M_p - I)}}$, where $M_p$ is the stability matrix, $\mu_p$ is the Maslov index, and $T_p^{\rm prim}$ is the primitive period \cite{Stockmann-1999,Haake-2019}. If the orbit involves multiple revolutions of a shorter orbit, the primitive period is defined to be the period of the latter orbit.
In the factors accounting for symmetries and spin, $g_p$ is the group element relating the initial and final points of $p$ when viewed in the full system as opposed to the fundamental domain, and $\chi_\alpha(g_p)$ is its character in the irrep $\alpha$. ${\rm tr}\, d_p$ is the trace of a matrix that on one hand depends on the precession of spin around the orbit. Crucially, this factor is now also affected by the presence of symmetries, and includes terms arising from the spin contributions to the quantum symmetry operators.

This paper is organised as follows. The necessary background on spin systems, symmetries and representation theory will be introduced in Section~\ref{sec: symmetries in spin systems}. We will then derive a Green function incorporating the effects of both symmetries and spin in Section~\ref{Sec: Symmetry reduced Green function}. The corresponding trace formula is obtained in Section~\ref{Sec: Symmetry reduced trace}. Finally, in Section~\ref{sec: spectral determinant} we extend our results to the spectral determinant, which is an important ingredient in semiclassical approaches to spectral statistics, before offering some concluding remarks in Section~\ref{sec: conclusions}.

\section{Symmetries in non-relativistic quantum systems with spin}

\label{sec: symmetries in spin systems}

In order to explain the derivation of the trace formula we must first provide a general background of how symmetries manifest in spin systems. We start with an overview of spin systems and then describe how (unitary) symmetries appear in such systems, and how they lead to the double group structure. The effects of the symmetries on these systems can then be explained using basic ideas of representation theory and their interaction with antiunitary (or time-reversal) operators.  Helpful references include \cite{Wigner-1959,Strange,Sakurai2,Tannoudji-2019,Hamermesh-1962,Lax-1974,Opechowski-1940,Batanoumy-2008,Doublegroups,Streitwolf-1971,Cornwell-1984,
Elliott-1979,Haake-2019}.

\subsection{Spin in non-relativistic quantum mechanics}\label{Spin systems}

In this paper we are interested in spin systems with arbitrary spin quantum number $S \in \{0, \frac{1}{2}, 1, \frac{3}{2}, \ldots\}$. The corresponding wave functions $\bpsi$ are $2S+1$ dimensional vectors, parameterised by the position $\bq$, and we assume for simplicity that the Hamiltonians are of so-called {\it Pauli type}, i.e.
\begin{equation}\label{Eqn: Pauli Hamiltonian}
\hat H =  \hat H_0 + \hat  \bS \cdot \hat{\bC}.
\end{equation}
Here $\hat H_0$ is a differential operator acting on the position argument
of the wavefunction, e.g.  $\hat H_0=\frac{{\hat\bp}^2}{2m}+V(\bq)$
where $\hat{\bp}=\frac{\hbar}{i}\nabla_\bq$ is the momentum operator. $\hat \bS$ is given by the $2S+1$ dimensional generalised Pauli matrices as $\hat \bS = \frac{\hbar}{2} \bsigma$, with $\bsigma =  (\sigma_x,\sigma_y,\sigma_z)$ in three-dimensional space, and $\hat{\bC}$ is a differential operator with $2S+1$ components.
If we follow a tensor notation (with spin degrees of freedom written before orbital degrees of freedom) and write out explicitly the sum over spin components, Eq.\ (\ref{Eqn: Pauli Hamiltonian}) turns into 
\begin{equation}  
\hat H   =  \Id_{2S+1} \otimes \hat{H}_0 + \sum_{a = x,y,z} \hat S_a \otimes \hat{C}_{a}
\end{equation}
where $\Id_{2S+1}$ is the $2S+1$ dimensional identity matrix.

 Possible choices for $\hat H_0$ and $\hat\bC$ are as follows \cite{Strange,Sakurai2,Tannoudji-2019}:

  \begin{itemize}

\item The Pauli equation is obtained from the leading non-relativistic limit of the Dirac equation.
For a particle of mass $m$ and charge $q$ in a magnetic field
$\bB({\bq})=\nabla\times\bA(\bq)$ and electric potential $\phi(\bq)$, it involves
\begin{align}  \label{H1}
\hat{H}_0&=\frac{1}{2m}\Bigg(\hat{\bp} -q \bA({\bq}) \Bigg)^2+q\phi(\bq),\quad\quad\quad
\hat\bC= -\frac{q }{m}  \bB({\bq}).
\end{align}
Here a term $mc^2$ (where $c$ is the speed of light) from the Dirac equation has been eliminated by a constant energy shift.
The Hamiltonian \eqref{H1} can alternatively be written as
\be \label{H1alt}
\hat H=\hat H_0+\hat\bS\cdot\hat\bC=\frac{1}{2m}\left[\bsigma\cdot\left(\hat\bp-q\bA(\bq)\right)\right]^2+q\phi(\bq).
\ee

\item
In the absence of a magnetic field, going to order $\frac{1}{c^2}$ in the expansion of the non-revalitivistic limit of the Dirac equation yields
\begin{align}\label{H2}
\hat{H}_0 &=  \frac{\hat{\bp}^2}{2m}+q\phi(\bq)-\frac{\hat{\bp}^4}{8m^3
c^2}+\frac{q^2\hbar^2}{8m^2c^2}\nabla^2\phi(\bq),
\notag\\ 
\hat\bC&=\frac{q}{2
m^2 c^2}      \nabla \phi({\bq})\times \hat{\bp} .
\end{align}

\item If we expand to order $\frac{1}{c^2}$ only in $\hat\bC$,  a particle in an electric potential $\phi(\bq)$   can be described by
\begin{align}  \label{H3}
\hat{H}_0&= \frac{\hat{\bp}^2}{2m}
+ q\phi({\bq}),\quad\quad\quad
 \hat\bC= \frac{q}{2 m^2 c^2}    \nabla\phi({\bq}) \times \hat{\bp}.
\end{align}

\end{itemize}

\subsection{Discrete symmetries  }\label{Symmetry groups}

We are interested in the case that our   Hamiltonian   commutes with a group of discrete unitary symmetry operators $\U$. We also have to consider antiunitary symmetries such as time-reversal invariance, but this will be postponed until
later.  We assume that each of the unitary symmetry operators can be written as a tensor product  
\be\label{U}
\U=\Us\otimes \Ug
\ee
where $\Us$ is a matrix acting in  $2S+1$ dimensional spin space. For $S=0$, $\Us$ amounts to multiplication with a phase factor that can  be disregarded as the wavefunction
is well defined only up to multiplication with such a factor, but for larger $S$ it plays a crucial role.  $\Ug$ acts on the orbital degrees of freedom through a geometrical operation $\R$ in position space, as in\footnote{The inverse of $\R$ is used by convention as with $\hat U_{{\rm geo},j}\psi(\bq)=\psi(\R_j^{-1}\bq)$, $j=1,2$
we then have $\hat U_{{\rm geo},1}\hat U_{{\rm geo},2}\,\psi(\bq)=\psi(\R_2^{-1}\R_1^{-1}\bq)=\psi((\R_1\R_2)^{-1}\bq)$, i.e., the quantum operators and the geometrical operations have the same behaviour under multiplication.}
\be\label{Ug}
\Ug\bpsi(\bq)=\bpsi(\R^{-1}\bq)
\ee
where $\bpsi(\bq)$ is $2S+1$ dimensional.
Altogether we thus have
 \begin{equation}\label{Eqn: Unitary form}
  \U \bpsi(\bq)  = \Us \bpsi(\R^{-1}\bq) ,
\end{equation}
and application of $\U$ to the Pauli Hamiltonian leads to
\begin{equation}\label{Eqn: Hamiltonian symmetry}
\U \left(\Id_{2s+1} \otimes \hat{H}_0+ \sum_{a } \hat S_a \otimes \hat{C}_a\right) = \Us  \otimes (\Ug  \hat H_0 )+ \U  \left( \sum_{a } \hat S_a \otimes \hat{C}_a\right).
\end{equation}
Using this result, one can show that $\U$ commutes with the Hamiltonian  under the following two conditions
\begin{enumerate}
\item \label{Item: Symmetry condition 1}$\Ug \hat{H}_0 = \hat{H}_0\Ug $
\item \label{Item: Symmetry condition 2}$\U   \left( \sum_{a } \hat S_a \otimes \hat{C}_a\right) = \left( \sum_{a } \hat S_a \otimes \hat{C}_a\right)\U $\end{enumerate}
i.e. $\hat H_0$ has to commute with the orbital part of the symmetry operator
and  $\hat \bS \otimes \hat{\bC}$ has to commute with the full operator.

\subsection{Rotations and double groups}

\label{sec:rotations}

In what follows we assume our geometrical symmetries act on position vectors $\bq \in \mathbb{R}^3$ and are described by $3 \times 3$ orthogonal matrices $\R \in O(3)$. Such matrices describe actions with the origin as a necessary fixed point and, hence, are referred to as point groups (see e.g.~\cite{Hamermesh-1962,Lax-1974}). Orthogonality implies $\det\R=\pm 1$ and, therefore, the matrices $\R$ can be classified into either \emph{proper rotations} about axes through the origin (i.e. $\det \R=1$) or \emph{improper rotations} (with $\det \R=-1$). The former comprise the group $SO(3)$, whereas the latter can be written as $\R= \I \mathcal{Q}$ with $\I\bq=-\bq$ the inversion operator and $\mathcal{Q} \in SO(3)$. Reflections are included in this formalism as well as they can be written as a combination of inversion with a rotation. Our results generalise also to translations, but since we need a finite position space this is only relevant for periodic boundary conditions.

\paragraph{Proper rotations in quantum mechanics.}
We now consider proper rotations. If $\R$ is  a proper (anticlockwise) rotation
by an angle $\theta$ about an axis through the origin with direction $ {\bn}$, the corresponding quantum operator $\Ug$ can be expressed in terms
of the angular momentum operator $\hat{\bL} = {\hat\bq} \times \hat{\bp}$
as 
\be
\Ug(\bn,\theta)=\exp\left(-\frac{i}{\hbar}\theta \hat\bL \cdot {\mathbf{n}}\right).
\ee
The operator $\Us$ for the corresponding spin rotation is given by \cite{Tannoudji-2019}
\be
\Us(\bn,\theta)=\exp\left(-\frac{i}{\hbar}\theta \hat \bS \cdot {\mathbf{n}}\right).
 \ee
 Note that this  spin rotation operator is also compatible with the form of the Pauli-type Hamiltonians.\footnote{
The above choice of $\Us(\bn,\theta)$ makes sure that $\hat\bS$ undergoes the same rotation that the gradient and hence $\hat\bp$ undergo under application of $\Ug(\bn,\theta)$. Assuming that $\phi$ is invariant under a given orbital rotation, the triple product arising in (\ref{H2}) and (\ref{H3}) for $\hat \bS\cdot\hat\bC$ is invariant under $\U(\bn,\theta)$ as all factors are rotated in the same way. If both $\phi$ and $\bB=\nabla\times \bA$ are invariant under orbital rotation, an analogous argument applies to
(\ref{H1}), noting that in this case $\bA$ also rotates like $\hat\bp$.}
 We thus obtain
\be
\U(\bn,\theta)=\Us(\bn,\theta)\otimes \Ug(\bn,\theta)=\exp\left(-\frac{i}{\hbar}\theta \hat\bJ \cdot {\mathbf{n}}\right)
\ee
where $\hat\bJ=\hat\bS+\hat \bL$ is the total angular momentum operator.
(In tensor notation we have $\hat{\bJ} = \hat \bS \otimes \Id + \Id \otimes \hat{\bL}$.)

\paragraph{Double groups.}
With the above definition we have
\be\label{twopi}
\Us(\bn,2\pi)=\begin{cases}\Id_{2S+1}&S=0,1,2,\dots\\
-\Id_{2S+1}&S=\frac{1}{2},\frac{3}{2},\frac{5}{2},\ldots
\end{cases}
\ee
i.e. a spin rotation by $2\pi$  flips the sign for half-integer spin.
Here we used that $\hat \bS\cdot\bn$ has eigenvalues given by $\hbar$ times an integer or half-integer, depending on whether $S$ is integer or half-integer. Eq.\ (\ref{twopi}) leads to

 \begin{equation}\label{Eqn: spin mapping}
\Us(\bn,\theta + 2\pi) = \Us(\bn,\theta)  \Us(\bn,2\pi)  = \left\{\begin{array}{ll}
\Us(\bn,\theta) & S = 0,1,2, \ldots \\
-\Us(\bn,\theta) & S = \frac{1}{2},\frac{3}{2},\frac{5}{2},  \ldots \end{array}\right.
\end{equation}
Hence, while for integer spin there is a unique spin matrix for each rotation
by $0\leq\theta<2\pi$ leaving the Hamiltonian invariant, for half-integer spin there are two such matrices,
\begin{equation}\label{doubling}
\Us(\bn,\theta)\quad\quad\text{and} \quad\quad \Us(\bn,\theta+2\pi)=-\Us(\bn,\theta).
\end{equation}
Considering $S=\frac{1}{2}$, the mathematical origin for this ambiguity is
that the rotation matrices are part of $SO(3)$ whereas the $S=\frac{1}{2}$
Pauli matrices generate $SU(2)$, and the natural mapping from $SO(3)$ to
$SU(2)$ is double valued.

This means that there are two quantum mechanical operators associated to
(proper) rotations, $\U(\bn,\theta)$ and $-\U(\bn,\theta)$. 
For  improper rotations there are also two operators, including  classical inversion as an additional
factor (see Appendix A).  

In an abstract notation we denote the sets of symmetry operations without minus signs
as $\Gamma$. This set contains the identity element $e$ for $\theta=0$. The elements with minus sign are written as $\bar eg$ with $g\in\Gamma$. Quantum mechanically
$\bar e$ amounts to a sign flip which for consistency has to be considered
as a separate group element. However, multiplications with other phase factors are not obtainable from rotations, and hence not included among the group elements. The overall group $G=\Gamma\cup\bar e\Gamma$
is doubled in size compared to the classical case, and hence called a double group \cite{Opechowski-1940,Streitwolf-1971,Doublegroups,Batanoumy-2008}.

\paragraph{Example: Double group of $C_3$.} The group $C_3$ consists of rotations by angles 0, $\frac{2\pi}{3}$ and $\frac{4\pi}{3}$. Thus choosing an axis, say the $z$-axis whose
direction is denoted by $\hat z$, it may be represented as $ \Gamma = C_3 = \{e,\R,\R^2\}$, where $\R$ denotes orbital rotation by $\frac{2\pi}{3}$ and $\R^3=e$ is the identity. However, if we identity $g$ with the quantum-mechanical rotation by $2\pi/3$ we have $\U(\hat z,2 \pi/3)^3 = \U(\hat z,2\pi) = -\Id_{2S+1}$ corresponding to $\bar{e}$ and $\U(\hat z,2 \pi/3)^6 = \Id_{2S+1}$ corresponding to $e$. This gives $G = C_6 = \{e,g,g^2,\bar{e}, \bar{e}g, \bar{e}g^2\}=C_3\cup \bar eC_3$ where $g^3 = \bar{e}$.

\subsection{Representation theory}

\paragraph{Irreducible representations and subspectra.}

Regardless of whether the symmetry group of the system is given by the original geometric group or the larger double group, the ideas of representation theory can be
applied to understand the system's spectral properties
(see e.g. \cite{Wigner-1959, Hamermesh-1962,Elliott-1979,Cornwell-1984,Batanoumy-2008}).

The starting point is that every element $g$ of our abstract symmetry group $G$ can be represented by a unitary operator $\U(g)$, see (\ref{U}), acting on the Hilbert space $\HS$. The
operators $\U(g)$ obey the same behaviour under multiplication
as the abstract group elements\begin{equation}\label{eqn: representation relation}
  \U(g_1)\U(g_2) = \U(g_1g_2) \quad  \forall g_1, g_2 \in G ,
\end{equation}
which is the defining property of a representation.
As symmetry operators, they commute with the Hamiltonian, $[\U(g),\hat{H}] = 0 \; \forall g \in G $.
One can show that this implies that that there exists a basis of the eigenfunctions of the Hamiltonian in which the operators $\U(g)$
are all block diagonal and the blocks are given by  irreducible  representations (irreps)
$\alpha$, that assign a matrix $\rho_\alpha(g)$ to each group element $g$. The $\rho_\alpha(g)$ are $s_\alpha \times s_\alpha$
unitary matrices, where $s_\alpha$ is the dimension of the irreducible representation $\alpha$ (irrep), and they satisfy the same composition relation
\begin{equation}
  \rho_{\alpha}(g_1)\rho_{\alpha}(g_2) = \rho_{\alpha}(g_1g_2) .
\end{equation}
Irreducibility means that the blocks cannot be split into smaller by choosing any other basis, and for finite symmetry groups $G$ there is only a finite number of different irreps
(up to unitary equivalence). In fact, $\sum_\alpha s_\alpha^2 = |G|$ where $|G|$ is the number of elements in $G$. Finally, the $s_\alpha$ eigenfunctions
corresponding to the same block all have the same energy eigenvalue.

We can thus label eigenfunctions by the irrep $\alpha$, an index $i=1\ldots s_\alpha$ that runs over the eigenfunctions for one block, and an index $n$
that labels different blocks corresponding to the same irrep $\alpha$, leading to the notation $\ket{\alpha,n,i}$. The action of a symmetry operator $\U(g)$ 
on a basis state is then given by
\begin{equation}\label{Eqn: Unitary action}
\U(g)\ket{\alpha,n,i} = \sum_{j=1}^{s_{\alpha}}[\rho_\alpha(g)]_{ji}\ket{\alpha,n,j}, \qquad \hat{H}  \ket{\alpha,n,j} = E^{(\alpha}_n \ket{\alpha,n,j} .
\end{equation}
One can see from this result that the spectrum of the Hamiltonian decomposes into subspectra associated to the different irreps $\alpha$. In an alternative
notation one combines the $s_\alpha$ eigenfunctions with the same $\alpha$ and $n$ into a tuple
\be
\ket{\alpha, n} = \begin{pmatrix} \ket{\alpha,n,1} \\ \vdots \\ \ket{\alpha,n,s_{\alpha}} \end{pmatrix} .
\ee
The action of a symmetry operator $\U(g)$ onto such a tuple is then given by 
\begin{equation}
\U(g)\ket{\alpha,n} = \rho_\alpha(g)^T\ket{\alpha,n} .
\end{equation}
Here the appearance of the transpose is a general feature of transformations of basis vectors.
In the following we will sometimes consider the Hilbert space $\HS_\alpha$  spanned by $|\alpha,n\rangle$, containing all 
$\ket{\Psi}$ that satisfy
\begin{equation}
\U(g)\ket{\Psi} = \rho_\alpha(g)^T\ket{\Psi}
\label{Eqn: U rho}.
\end{equation}

Finally, in the usual Hilbert space of wavefunctions, the projection operator onto a subspace associated to a given irrep $\alpha$ is given by
\begin{equation}\label{projector}
\hat P_{\alpha} = \frac{s_{\alpha}}{|G|}\sum_{g \in G }\chi_{\alpha}(g) \U^{\dagger}(g) \end{equation}
where $\chi_\alpha(g)=\tr\rho_\alpha(g)$ is called the character of the group element $g$ in the representation $\alpha$.

\paragraph{Types of representations.}

All unitary irreps of a finite group may be classified into one of three
types,
depending on whether the corresponding matrices can be brought to real or
quaternion-real (pseudo-real) form through a suitable similarity transformation,
or not. Firstly, if there does not exist a unitary $Z$ such that 
\begin{equation}\label{Eqn: Irrep classification}
{\rho_{\alpha}(g)}^{*}=Z \rho_{\alpha}(g)Z^{-1}, \quad \forall g \in G,
\end{equation}
where $^*$ denotes complex conjugation, then $\alpha$ cannot be brought to
such a form and is said to be \emph{complex}.
On the other hand, if such an $Z$ exists then either $ZZ^* = \Id_{s_{\alpha}}$,
in which case $\alpha$ is \emph{real}, or $ZZ^* = -\Id_{s_{\alpha}}$, corresponding
to a \emph{pseudo-real} irrep.  It can be shown that the latter property implies that $s_{\alpha}$
must be even for   pseudo-real $\alpha$.

By using (\ref{Eqn: Irrep classification}) and the following orthogonality
relation for the characters
\begin{equation}\label{orthog}
\frac{1}{|G|}\sum_{g \in G}\chi_{\alpha}(g) \chi^{*}_{\beta}(g)=\delta_{\alpha
\beta}.
\end{equation}
one can show that the type of the irrep $\alpha$ can be read off from  the Frobenius-Schur indicator \cite{Frobenius-Schur}
\begin{equation}\label{Eqn: FS indicator}
\FS_\alpha : = \frac{1}{|G|}\sum_{g \in G} \chi_{\alpha}(g^2) =  \left\{
\begin{array}{cl}
0 & \quad \alpha \ \mbox{complex} \\
+1 & \quad \alpha \ \mbox{real} \\
-1 & \quad \alpha \ \mbox{pseudo-real} . \end{array}\right.
\end{equation}

\paragraph{Irreducible representations of double groups.}

For double groups, the irreps may be further classified into one of two forms - due to the nature of the group element $\bar{e}$. Since $\bar{e}$ commutes with every element of the symmetry group one can show (using Schur's lemma \cite{Cornwell-1984}) that $\rho_{\alpha}(\bar{e}) = \kappa \Id_{s_{\alpha}}$. In addition, $\rho(\bar{e})^2 = \rho(e) = \Id_{s_{\alpha}}$ implies that $\kappa = \pm 1$. In the case  $\kappa = 1$ the group elements $g$ and $\bar{e}g$ therefore have the same representation and hence form a representation of $\Gamma$ when restricted to that subgroup. These representions are referred to as \emph{standard} irreps. Alternatively, when $\kappa = -1$, $g$ and $\bar{e}g$ are represented by $\rho_\alpha(g)$ and $\rho_\alpha(\bar{e}g)=\rho_\alpha(\bar{e})\rho_\alpha(g)=-\rho_\alpha(g)$. The corresponding representations are called the \emph{extra} irreps. (In fact all these irreps are known as \emph{projective representations} of the geometric symmetry group $\Gamma$, although we shall not go into detail on this point.)

Importantly, due to the nature of our unitary operators $\U(g)$, the only irreps that appear in the decomposition of the Hilbert space $\HS$ are the extra irreps. This can be deduced by investigating the form of the operator $P_{\alpha}$ that projects onto to the subspace $\alpha$. If we split the
sum over group elements into geometric group elements and elements containing
a factor $\bar e$, Eq.\ (\ref{projector}) turns into    
\begin{equation}\label{Eqn: Projection operator}
\hat P_{\alpha} = \frac{s_{\alpha}}{|G|}\sum_{g \in G }\chi_{\alpha}(g) \U^{\dagger}(g) =  \frac{s_{\alpha}}{|G|}\sum_{g \in \Gamma }\Big(\chi_{\alpha}(g) \U^{\dagger}(g)+\chi_{\alpha}(\bar{e}g) \U^{\dagger}(\bar{e}g)\Big).
\end{equation}
However, by the nature of the matrices $\rho_{\alpha}$, for $g \in \Gamma$ we have $\chi_{\alpha}(\bar{e}g) = \kappa \chi_{\alpha}(g) $. Therefore we obtain
\begin{equation}\label{projection_double}
\hat P_{\alpha} = \frac{s_{\alpha}}{|G|}\sum_{g \in \Gamma }[\chi_{\alpha}(g) - \kappa\chi_{\alpha}(g)] \U^{\dagger}(g) = 
\left\{\begin{array}{ll}
0 & \quad \kappa=1\\
\frac{s_{\alpha}}{|\Gamma|}\sum_{g \in \Gamma }\chi_{\alpha}(g) \U^{\dagger}(g)& \quad \kappa=-1 ,
\end{array}\right.
\end{equation}
where we   also used $\U(\bar{e}g) = -\U(g) $. As the projection on the space spanned by the eigenfunctions related to each standard irrep vanishes, we can conclude that there are no such eigenfunctions and   thus limit   our interest to the extra irreps.
  
The Frobenius-Schur indicator for these irreps can conveniently be written in terms of the elements of the geometric symmetry group $\Gamma$ as  
\begin{equation}
\FS_{\alpha} =\frac{1}{|G|}\sum_{g \in G} \chi_{\alpha}(g^2)= \frac{1}{2|\Gamma|}\sum_{g \in \Gamma}[ \chi_{\alpha}(g^2)
+ \chi_{\alpha}((\bar{e}g)^2) ] = \frac{1}{|\Gamma|}\sum_{g \in \Gamma} \chi_{\alpha}(g^2)
\end{equation}
where we used that   $\bar{e}$ commutes with every group element and that
$\bar{  e}^2 = e$.

\paragraph{Example: Double group of $C_3$.} As we have established, the double
group of $C_3$ is given by $C_6$. This group   has 6   irreducible
representations $\alpha$ which are one-dimensional and hence coincide with  the characters
\be
\chi_{\alpha}(g^r) = e^{\pi i\alpha r/3}, \quad \alpha,r = 0,1,\ldots, 5.
\ee The \emph{standard} irreps are given by $\alpha=2\beta=0,2,4$
which leads to $\chi_{2\beta}(g^r) = e^{2\pi i r \beta/3}$  coinciding
with the irreps of $C_3$ when $r=0,1,2$. The \emph{extra} irreps occur for
$\alpha=1,3,5$. Checking the Frobenius-Schur indicator we find
\be
\FS_{\alpha} =  \frac{1}{3}\sum_{g \in C_3} \chi_{\alpha}(g^2) = \frac{1}{3}
\sum_{r=0}^2 e^{2 \pi i \alpha r /3} = \left\{\begin{array}{ll} 0 & \alpha=1
\\
1 & \alpha =3 \\
0 & \alpha = 5 \end{array}\right.
\ee
Hence the extra irreps $\alpha = 1,5$ are complex and $\alpha = 3$ is real.

\subsection{Antiunitary symmetries}\label{Sec: Antiunitary symmetries}

In addition to the unitary symmetries outlined above, 
quantum systems may also have an antiunitary symmetry, alternatively referred to as \emph{generalised time-reversal symmetry} \cite{Lax-1974,Streitwolf-1971,Batanoumy-2008,Haake-2019}. In addition to commuting with
the Hamiltonian an antiunitary symmetry operator $\TR$ has to satisfy
\be
\langle\TR\psi_1|\TR\psi_2\rangle=\langle\psi_1|\psi_2\rangle^*
\ee
as well as the antilinearity property
\be
\TR(c_1\psi_1+c_2\psi_2)=c_1^*\TR\psi_1+c_2^*\TR\psi_2
\ee
for all complex numbers $c_1$, $c_2$ and wavefunctions $\psi_1$, $\psi_2$.

For spinless systems the most common antiunitary symmetry is the conventional
time-reversal symmetry $\TR=\K$, where  $\K$ is the complex conjugation operator. In particular this operator satisfies
\begin{equation}\label{Eqn: TR properties}
\TR \hat\bq \TR^{-1} = \hat\bq, \hspace{10pt} \TR \hat\bp \TR^{-1} = -\hat\bp, \hspace{10pt} \TR^{-1} \hat\bL \TR^{-1} = -\hat\bL,
\end{equation}
as expected from time-reversing the motion on classical trajectories.
For spin systems, a natural definition for a conventional time-reversal operator
is 
\begin{equation}\label{Eqn: TR operator}
\TR  =  (e^{i\pi \hat S_y/\hbar} \otimes \Id) \K,
\end{equation}
which satisfies (\ref{Eqn: TR properties}) with $\hat\bL$ replaced by $\hat\bJ$
(as complex conjugation flips the sign of $\hat S_y$, and conjugation with $\hat S_y$ flips the sign of $\hat S_x$ and $\hat S_z$).
This operator squares to  
\be
\TR^2 = (e^{i 2 \pi \hat S_y/\hbar} \otimes \Id) =  \pm \Id
\ee
respectively for integer and half-integer spin (and hence integer and half-integer eigenvalues of $\hat S_y$). 

\subsubsection{Generalized time-reversal invariance.}
Among the Pauli-type Hamiltonians, (\ref{H2}) and (\ref{H3}) are invariant under the conventional time reversal operator (\ref{Eqn: TR operator}), 
as it takes $\hat\bp\to-\hat\bp$ and $\hat \bS\to -\hat \bS$. For (\ref{H1}) this is not the case, unless the vector potential and hence the magnetic field vanishes. However, depending on the form of $\phi$, $\bA$ and $\bB$, there may be a different antiunitary symmetry, including an additional spin operator $\Us$ and/or geometrical operator $\Ug$,
\be
  \TR  =  (\Us e^{i\pi \hat S_y/\hbar} \otimes \Ug) \K.
\ee
For example, one can show that if a geometrical rotation by $\pi$ leaves the electric potential $\phi$ invariant and flips the sign of $\bB$, this leads to an antiunitary symmetry with the above rotation as $\Ug$ and $\Us=\Id$.

As the product of any given antiunitary $\TR$ with a unitary symmetry is also an antiunitary
symmetry, the full group of all symmetry operations is  of the form 
\begin{equation}
{G}_{\rm full} = G \cup \TR G.
\end{equation}
Groups $G_{\rm full}$ in which  an antiunitary operator such
 as $\TR$ in (\ref{Eqn: TR operator}) commutes with every unitary operator are called \emph{factorisable}. Otherwise, the group is \emph{non-factorisable}, which may occur if the Hamiltonian is not invariant under conventional time reversal, but only under its combinations with  a unitary transformation that does not commute with the unitary
symmetries of the system.

We are primarily interested in systems where, if present, the time-reversal operator $\TR$ squares to $\pm\Id$ and $G_{\rm full}$ is factorisable.
In general, a description of systems with symmetries where this condition is not met requires the use of corepresentations. These are extensions of representations to symmetry groups that in addition to unitary operators also contain antiunitary operators. The latter are then represented by products of unitary matrices with the complex conjugation operator.
 Corepresentations also play a crucial role in Dyson's seminal paper \cite{Dyson-1962} which studies, within random-matrix theory, how time-reversal properties and symmetries affect the choice of an appropriate random-matrix emsemble. We will further comment on the connection to Dyson's work when using the present results to semiclassically determine spectral statistics in \cite{Blatzios-2024b}. Corepresentations have so far not been incorporated in a semiclassical approach. While the restrictions on $\TR$ are relevant for \cite{Blatzios-2024b}, in the present derivation of a trace formula, time-reversal properties and hence the conditions for $\TR$
only affect the treatment of Kramer's degeneracy below.
In a different context, a contribution to the theory of corepresentations, based on \cite{Winter-2017}, is in preparation.

As shown in Appendix B, for $\TR^2=\pm \Id$ it is justified to consider only $\Us=\Id$, i.e., generalised time-reversal operators are only distinguished from the conventional one by an additional geometrical operation $\Ug$.
As $\Ug$ is real, it   leads to a factor $\Ug^2$ in $\TR^2$, which given the choice between $\Id$ and $-\Id$ can only be equal to $\Id$. Hence, 
$\TR^2$ behaves as for conventional time-reversal invariance and is equal to $\Id$ for integer $S$ and $-\Id$ for half-integer $S$.

\paragraph{Kramer's degeneracy.}

If the  time reversal operator satisfies $\TR^2 = -\Id$ then we obtain a 2-fold degeneracy in the system known as \emph{Kramer's degeneracy}. This is easily verified since
\begin{equation}\label{Eqn: Kramers proof}
\langle \psi | \TR \psi \rangle = \langle \TR \psi | \TR^2  \psi \rangle^*  =  -\langle \TR \psi | \psi \rangle^* = -\langle \psi | \TR \psi \rangle
\end{equation}
implies that the state $\ket{\psi}$ and its time-reversed partner $\ket{\TR \psi}$ are orthogonal. When no unitary symmetries are present the situation is clear - every energy level appears with a doubled multiplicity. However, in the presence of unitary symmetries the situation is more delicate and depends on the type of irreducible representation. This case will be dealt with below.

\paragraph{Transferred time-reversal operator.}

If $\TR$ is a symmetry of the Hamiltonian, this does not necessarily mean that it is also a symmetry in the subspace $\HS_\alpha$ associated to an irrep $\alpha$.
If one applies $\TR$ to  $\ket{\alpha,n} \in \HS_\alpha$ then the result $\TR \ket{\alpha,n} $ may no longer be an element of $\HS_\alpha$.
This is because due to the antiunitary nature of the time-reversal operator, $\TR\ket{\alpha,n}$ satisfies (\ref{Eqn: U rho}) with the complex conjugate of $\rho_\alpha(g)$, which may or may not be unitarily equivalent to $\rho_{\alpha}(g)$ (see Eqn. (\ref{Eqn: Irrep classification})).
To obtain a symmetry in the subspace $\HS_\alpha$ we must instead consider the \emph{transferred} time-reversal operator \cite{Zirnbauer-2011}
\begin{equation}
\TR_{\alpha} = Z \otimes \TR,
\end{equation}
where $Z$ is the matrix of dimension $s_\alpha$ given in (\ref{Eqn: Irrep classification}), and the tensor product is taken between
the corresponding space and the space containing both spin and orbital degrees of freedom. 
With this definition one can show that $\TR_\alpha$ commutes with the symmetry-reduced Hamiltonian in the subspace $H_\alpha$,
and crucially  $\TR_{\alpha} \ket{\alpha,n}$ satisfies (\ref{Eqn: U rho}), provided that a suitable $Z$ exists, i.e.\ that $\alpha$ is either real or pseudo-real. 

Interestingly, the transferred time-reversal operator squares to
\begin{equation}\label{Eqn: Transferred TR operator squared}
\TR_{\alpha}^2 = ZZ^* \otimes \TR^2 = \left\{\begin{array}{r l}  \Id\otimes\TR^2 & \quad \alpha \ \mbox{real} \\
-\Id\otimes\TR^2 & \quad \alpha \  \mbox{pseudo-real} \end{array} \right.
\end{equation}
i.e. it squares to the same values as $\TR^2$ for real irreps, but the sign is flipped for pseudo-real irreps. This suggest a swapping between GOE and GSE type behaviour for pseudo-real irreps, which we will  explain from a semiclassical point of view in \cite{Blatzios-2024b}. 

\paragraph{Kramer's degeneracy in the presence of unitary symmetries.}
We note that, as $\TR_\alpha$ commutes with the Hamiltonian $\hat{H}_\alpha$, both $|\alpha,n\rangle$ and $\TR_{\alpha} \ket{\alpha,n}$
are eigenstates of $\hat{H}_\alpha$ with the same eigenvalue. Thus, if $\TR^2 = \Id$, i.e. integer spin, we find from (\ref{Eqn: Transferred TR operator squared}) that Kramer's degeneracy exists in those subspaces $\HS_{\alpha}$ for which $\alpha$ is pseudo-real. In contrast, when $\TR^2 = -\Id$ then (\ref{Eqn: Transferred TR operator squared}) implies we have a Kramer's degeneracy when $\alpha$ is real. At first sight, this might suggest that the Kramer's degeneracy in full space (implied by $\TR^2 = -1$) disappears for pseudo-real subspaces, however in this case it is absorbed into the $s_\alpha$-degeneracy which is even for pseudo-real irreps.

\section{Green function for spin systems with symmetries}\label{Sec: Symmetry reduced Green function}

To semiclassically study the spectral statistics of spin systems with additional symmetries, we need a trace formula applicable to such systems. In the following we will derive such a trace formula by combining the ideas of Bolte and Keppeler
\cite{Bolte-1999, Keppeler-2003}, who derived a trace formula for spin systems, with those of Robbins \cite{Robbins-1989} (see also Lauritzen \cite{Lauritzen-1991}), who considered the effect of symmetries for spinless systems.  
We will start from a semiclassical approximation for the Green function of the full spin system, and then determine the resulting Green function for a subspectrum. The latter will, in the following section, allow us to obtain a level density for a subspectrum.

\subsection{Green function for a spin system}

The spectrum of a quantum system can be obtained from its Green function
\begin{equation}
G(\bq,\bq_0,E)=\langle \bq|(E+i\eta-\hat H)^{-1}|\bq_0\rangle 
\end{equation}
where $i\eta$ represents a positive imaginary part whose size is taken to zero. For spin systems $G(\bq,\bq_0,E)$ is matrix-valued due to the spin degrees of freedom.
A semiclassical approximation for the Green function of a spin system was derived by Bolte and Keppeler \cite{Bolte-1999,Keppeler-2003}  in the form
\begin{equation}\label{Eqn: Full Greens function}
G(\bq,\bq_0,E)\approx\sum_{\gamma: \bq_0 \rightarrow \bq} d_{\gamma} D_{\gamma} e^{i S_{\gamma}( E)/\hbar}
\end{equation}
which, like $G(\bq,\bq_0,E)$, is matrix-valued.
Here the sum is taken over all classical trajectories from $\bq_0$ to $\bq$ (governed by the classical Hamiltonian $H_0$) with energy $E$. $S_\gamma=\int_\gamma\bp\cdot d\bq$
is the associated classical action and $D_{\gamma}$ is the usual stability amplitude (see e.g. \cite{Haake-2019,Stockmann-1999})
\begin{equation}
D_{\gamma}=\frac{e^{-i \mu_{\gamma}\frac{\pi}{2}}}{i \hbar (2 \pi i \hbar)^{(f-1)/2}}\sqrt{\left|\text{det}\begin{pmatrix}
\frac{\partial^2 S}{\partial \bq \partial \bq'} & \frac{\partial^2 S}{\partial \bq \partial E}\\
\frac{\partial^2 S}{\partial \bq' \partial E}& \frac{\partial^2 S}{\partial E^2}
\end{pmatrix}\right|},
\end{equation}
involving derivatives of the action as well as  the Maslov index $\mu_\gamma$. Here $f$ is the number of degrees of freedom. The new feature, in comparison to spinless systems, is the $(2S +1) \times (2S +1)$ matrix valued amplitude $d_{\gamma}$ that describes the spin precession along $\gamma$. 
More specifically, this matrix is the solution of the first order differential equation 
\begin{equation}\label{Eqn: Spin precession}
\dot{d}_{\gamma}(t)  = -\frac{i}{2} \hat \bS \cdot \bC(\mathbf{q}_{\gamma}(t),\mathbf{p}_{\gamma}(t)) d_{\gamma}(t),
\end{equation} 
subject to the initial condition 
\begin{equation} \label{Eqn: Spin initial condition}
d_{\gamma}(0)=\mathbf{1}_{2s+1}.
\end{equation}
Here $\bq_{\gamma}(t)$ and $\bp_{\gamma}(t)$ denote the position and momentum of the particle along $\gamma$ after a time $t$ and $\bC(\bq,\bp)$ is the Weyl symbol corresponding to the operator $\hat{\bC}$ in (\ref{Eqn: Pauli Hamiltonian}). This symbol can be obtained from $\hat \bC$  if we write $\hat\bC$ in Weyl ordered form and replace position and momentum operators by the position and momentum variables $\bq$ and $\bp$. Equivalently we can use the formula
\be
\bC(\bq,\bp) := \int_{\mathbb{R}^n} d\by e^{-i\bp \cdot \by/\hbar} \bra{\bq + \by/2}{\hat \bC}\ket{\bq - \by/2}\;.
\ee
In this semiclassical picture the classical motion remains unaffected by the quantum spin. On the other hand the spin is driven by the classical motion. Following \cite{Bolte-1999,Keppeler-2003,Muller-2005} we assume the classical motion is fully chaotic, which in turn leads to ergodicity in the spin space $SU(2)$. 

\subsection{Green function for a subspectrum}

\label{subgreen}

The Green function of a subspectrum can be obtained if we apply the projection operator \begin{equation}\label{Eqn: projector_again}
P_{\alpha} = \frac{s_{\alpha}}{|G|}\sum_{g \in G }\chi_{\alpha}(g) \U^{\dagger}(g) \end{equation}
from  (\ref{projector})
to  the $G(\bq,\bq_0,E)$ from (\ref{Eqn: Full Greens function}). We consider symmetry operators of the form  $\U(g)=\Us(g)\otimes \Ug(g)$ as in (\ref{U}) and assume that all operators in our symmetry group have a nontrivial geometrical component. This excludes double groups to which we will return later. We can now use the same symbol for the abstract group element and for the operation in position space, and hence write 
\be
\Ug(g)^\dagger \psi(\bq)=\psi(g\bq),
\ee
see (\ref{Ug}). Application of $P_\alpha$ then leads to  
\begin{align}\label{Galphapre}
G_\alpha(\bq,\bq_0,E)=P_{\alpha}G(\bq,\bq_0,E)&\approx\frac{s_\alpha}{|G|}\sum_{g\in G}\chi_\alpha(g)\Us(g)^\dagger G(g\bq,\bq_0,E) \notag\\
&=\frac{s_\alpha}{|G|}\sum_{g\in G}\sum_{\gamma: \bq_0 \rightarrow g\bq} \chi_\alpha(g)\Us(g)^{\dagger}d_{\gamma} D_{\gamma} e^{i S_{\gamma}( E)/\hbar}.
\end{align}

We now apply the methodology of Robbins \cite{Robbins-1989} and  rewrite this Green function in terms of trajectories in a fundamental domain. A fundamental domain is a part of position space from which all other parts of the system can be accessed by symmetry operations, such that they can be seen as copies of the fundamental domain. 
The different copies of the fundamental domain may not overlap, except at boundaries\footnote{Trajectories along the boundary must be dealt with appropriately and we omit their contributions in this present context.}. 
We define a dynamics inside the fundamental domain such that trajectories leaving the fundamental domain are 'folded back'  into the original copy of the domain
using the relevant symmetry operation. This means that a trajectory reaching the boundary reemerges from a point on the boundary that is related to the previous
one by a symmetry operation. Otherwise, the dynamics within the fundamental domain are the same as for the full system.

Consequently, the trajectories $\bar\gamma$ in the fundamental domain (not confined to the boundary) are in one-to-one correspondence with trajectories $\gamma$ in the full system, with identical actions and stability amplitudes. This follows from \cite{Robbins-1989} without modification, as the classical dynamics of our system is just given by the spinfree Hamiltonian $\hat H_0$. The sum in (\ref{Galphapre}) can thus be written in terms of trajectories in the fundamental domain. 
As we will eventually identify $\bq$ and $\bq_0$, we are only interested in the situation that both points are located in the same copy of the fundamental domain.
Then the sum in (\ref{Galphapre}) can be written as a sum over all trajectories in the fundamental domain ending at $\bq$, which is the counterpart of all end points $g\bq$ in (\ref{Galphapre}). This gives
\begin{equation}\label{Galpha}
G_\alpha(\bq,\bq_0,E) \approx \frac{s_{\alpha}}{|G|}\sum_{\bar{\gamma}: {\bq}_0 \to {\bq}}\chi_{\alpha}(g_{\bar{\gamma}})\underbrace{\Us(g_{\bar\gamma})^\dagger d_\gamma}_{=:d_{\bar{\gamma}}} D_{\bar{\gamma}}e^{iS_{\bar{\gamma}}(E)/\hbar}
\end{equation}
where the group element is written $g_{\bar \gamma}$ instead of $g_\gamma$ as it  can be deduced purely from the parts of the boundary hit by $\bar\gamma$. 

\subsection{Spin factor}

To finalise our interpretation of (\ref{Galpha}) we express  the spin factor $d_{\bar{\gamma}}=\Us(g_{\bar\gamma})^\dagger d_\gamma$ in terms of the dynamics in the fundamental domain. We will ultimately obtain an expression for $d_{\bar\gamma}$ that involves a product of spin factors associated to subsequent parts of trajectory in different copies of the fundamental domain, interspersed with factors associated to the motion from one copy to the next.

While the trajectory remains in the same copy of the fundamental domain $\Us(g_{\bar\gamma})^\dagger$ is constant and $d_{\bar{\gamma}}=\Us(g_{\bar\gamma})^\dagger d_\gamma$ obeys the equation
\begin{align}
\dot d_{\bar\gamma}(t)&=\Us(g_{\bar\gamma})^\dagger\dot d_\gamma(t)\nonumber\\
&=-\frac{i}{2}\Us(g_{\bar\gamma})^\dagger   \hat \bS\cdot \bC(\bq_\gamma(t),\bp_\gamma(t))d_\gamma(t)\nonumber\\
&=-\frac{i}{2}\Us(g_{\bar\gamma})^\dagger   \hat \bS\cdot \bC(\bq_\gamma(t),\bp_\gamma(t))\Us(g_{\bar\gamma}) d_{\bar\gamma}(t)\nonumber\\
&=-\frac{i}{2}   \hat \bS\cdot\Big(\Ug(g_{\bar\gamma}) \bC(\bq_\gamma(t),\bp_\gamma(t))\Ug(g_{\bar\gamma})^\dagger\Big) d_{\bar\gamma}(t)
\end{align}
where we used that $\hat \bS\cdot\bC$ is invariant under $\U(g_{\bar \gamma})=
\Us(g_{\bar \gamma})\otimes \Ug(g_{\bar \gamma})$ and hence conjugation with $\Us$ is equivalent to conjugation with $\Ug^\dagger$. The latter conjugation then applies the inverse geometric symmetry operation to  $\bC$. As this leads back to the fundamental domain, we obtain  
\begin{equation}
\label{diff1}
\dot d_{\bar\gamma} (t)=-\frac{i}{2} \hat \bS\cdot\bC(\bq_{\bar \gamma}(t),\bp_{\bar \gamma}(t))d_{\bar \gamma}(t)
\end{equation}
where $\bq_{\bar\gamma}=g^{-1}\bq_\gamma$, $\bp_{\bar\gamma}=g^{-1}\bp_\gamma$. We thus obtain the immediate analogue of Eq.\ (\ref{Eqn: Spin precession}) for the dynamics inside the fundamental domain.

When the trajectory crosses the border of the fundamental domain from a copy connected to the group element $g_{j-1}$ to a copy associated to $g_j=h_jg_{j-1}$ the factor $\Us(g_{\bar\gamma})^\dagger$ in $d_{\bar\gamma}$ changes from $\Us(g_{j-1})^\dagger$ to $\Us(g_j)^\dagger$ which amounts to left multiplication with $\Us(h_j)^\dagger$. To write $d_{\bar\gamma}$ explicitly we thus split our trajectory into segments $\bar\gamma_j$ ($j=0\ldots n$) in between crossings to a different copy of the fundamental domain. We thus obtain an alternating product 
\begin{equation}\label{Eqn: Spin precession FD}
d_{\bar{\gamma}} = d_{\bar{\gamma}_n}\Us(h_n)^{\dagger} d_{\bar{\gamma}_{n-1}}\Us(h_{n-1})^{\dagger}\ldots d_{\bar{\gamma}_1} \Us(h_1)^{\dagger} d_{\bar{\gamma}_0}.
\end{equation}
Here $d_{\bar\gamma_j}$ evolves according to (\ref{diff1}) with the initial condition $\Id$.

We thus see that $d_{\bar\gamma}$ evolves  like $d_\gamma$, however depending on the trajectory inside the fundamental domain and with additional factors related to the spin components of the symmetry operations, arising whenever the trajectory hits the boundary of the fundamental domain.

\section{Trace formula for spin systems with symmetries}\label{Sec: Symmetry reduced trace}


In the standard derivation of the Gutzwiller trace formula, the level density $\rho(E)=\sum_j(E-E_j)$ is accessed from the Green function via
\begin{equation}
\rho(E)  = -\frac{1}{\pi} \lim_{\epsilon \to 0} \Im  \int d  \bq \  G(\bq,\bq,E+i\epsilon)
\end{equation}
where the integral over $\bq$ is over the configuration space and can be interpreted as a trace.
The counterpart for our problem, i.e., a subspectrum of a spin system, is
\begin{equation}\label{rhofromG}
\rho_{\alpha}(E)  = -\frac{1}{\pi s_{\alpha}} \lim_{\epsilon \to 0} \Im  \int d  \bq \ \tr G_\alpha(\bq,\bq,E+i\epsilon)
\end{equation}
where we divide out the degeneracy $s_\alpha$ that the energy levels of our subspectrum have due to the geometrical symmetry. However, we do not correct for Kramer's degeneracy if it applies to our subspectrum. An additional trace,
denoted by ${\rm tr}$, is taken over the $(2S+1)\times(2S+1)$ spin matrices as in \cite{Bolte-1998, Bolte-1999a}.
The integral in \eqref{rhofromG} is over the complete original symmetric system. 
Note that we assume that the energy spectrum of the Pauli-type Hamiltonians in (\ref{Eqn: Pauli Hamiltonian}) that we consider is discrete. Conditions under which this holds
are discussed in~\cite{Bolte-2000}.

Now, inserting the Green function (\ref{Galpha}) into (\ref{rhofromG}) we obtain
\begin{equation}
\rho_{\alpha}(E) \approx -\frac{1}{\pi} \lim_{\epsilon \to 0}   \Im\int d \bq \frac{1}{|G|}    \ \sum_{\bar{\gamma}: {\bq} \to {\bq}}\chi_{\alpha}(g_{\bar{\gamma}})\tr (d_{\bar{\gamma}})\,D_{\bar{\gamma}}  e^{iS_{\bar{\gamma}}(E)/\hbar} \, .
\end{equation}
Here $\bq$ are points in any of the $|G|$ copies of the fundamental domain, and then $\bar\gamma$ denotes closed paths that are restricted to this copy by the dynamics that was in introduced in section \ref{subgreen}.  As the paths in these $|G|$ different copies all give the same contributions, we can integrate only over one copy of the fundamental domain and multiply the result by $|G|$. This gives  
\begin{equation}
\rho_{\alpha}(E) \approx -\frac{1}{\pi} \lim_{\epsilon \to 0}   \Im\int_{\rm F.D.} d \bq      \ \sum_{\bar{\gamma}: {\bq} \to {\bq}}\chi_{\alpha}(g_{\bar{\gamma}}) \tr (d_{\bar{\gamma}})\,D_{\bar{\gamma}}  \,e^{iS_{\bar{\gamma}}(E)/\hbar} .
\end{equation}
This agrees with standard derivation of Gutzwiller's trace formula, except for the factor $\chi_{\alpha}(g_{\bar{\gamma}})$ accounting for symmetry, 
the factor  $\tr d_{\bar{\gamma}}$ accounting for spin, and the dynamics being restricted to the fundamental domain. If we evaluate the integral as in the usual derivation according to Gutzwiller \cite{Gutzwiller-1970,Gutzwiller-1971} (see also \cite{Stockmann-1999,Haake-2019} for textbooks) we arrive at 
\begin{equation}\label{Eqn: Semiclassical dos}
\rho_{\alpha}(E) \approx \bar{\rho}_{\alpha}(E) + \frac{1}{\pi \hbar}\Re \sum_{p}\chi_{\alpha}(g_p)\tr(d_p) \,A_{p} e^{iS_p(E)/\hbar}.
\end{equation}
We postpone the discussion of the first term to the next section. The sum in the second term is over the classical periodic orbits $p$ in the fundamental domain. The amplitude factor $A_p$ has the form
\begin{equation}
A_p = \frac{T_p^\text{prim} e^{-i\mu_p \pi/2}}{\sqrt{|\det(M_p - I)}|},
\end{equation}
where $M_p$ is the stability matrix of the periodic orbit $p$, $\mu_p$ is its Maslov index and $T_p^{\text{prim}}$ is its primitive period.  

Note that the derivation of \eqref{Eqn: Semiclassical dos} used that the quantities $\chi_{\alpha}(g_p)$ and $\tr(d_p)$ are constant along a periodic orbit and do not depend on the choice of a starting point on this orbit. For $\tr(d_p)$ this follows from the fact that a different starting point corresponds to a cyclic permutation of the product of matrices in (\ref{Eqn: Spin precession FD}), and therefore the trace $\tr(d_p)$ remains invariant. Similarly, we can decompose $g_p$ into a product of generators of $G$ associated to the boundary on the fundamental domain and then the character $\chi_{\alpha}(g_p)$ is also invariant under such a cyclic permutation.

\subsection{Weyl term}

\label{Weyl}

In (\ref{Eqn: Semiclassical dos}), $\bar{\rho}_{\alpha}(E)$ is the semiclassical  approximation to the mean density of states obtained by integrating over all paths of zero length for which the stationary-phase approximation is not applicable. For comparison, in systems without symmetries and without spin the contribution of these zero-length trajectories to the mean density of states is asymptotically given by
\be
\bar{\rho}_0(E)  \sim \frac{|\Omega(E)|}{(2\pi\hbar)^{f}} ,
\ee
in the semiclassical limit $\hbar \to 0$. Here $f$ is the number of degrees of freedom and $|\Omega(E)|$ is the volume of the energy shell at energy $E$.

However, in comparison to the spinless case without symmetries, $G_{\alpha} (\bq,\bq ,E)$ also involves  a factor $\frac{1}{|G|}$ from (\ref{Galpha}) as well as a factor
$\chi_\alpha(e)=s_\alpha$ since for zero-length  trajectories the group element relating the initial and final copies of the fundamental domain is $e$. Furthermore, spin brings about a factor $\Id_{2S+1}$ which after taking the trace yields   $2S+1$. Altogether we thus have
\be\label{rhoav}
\bar{\rho}_{\alpha}(E) \sim \frac{ s_{\alpha}(2S+1)|\Omega(E)|}{  |G|(2 \pi \hbar)^f} 
\ee
which combines modifications from \cite{Robbins-1989} and \cite{Bolte-1998, Bolte-1999a}. 

The mean density of the original symmetric system
\be
\bar{\rho}(E) \sim \frac{ (2S+1)|\Omega(E)|}{  (2 \pi \hbar)^f} 
\ee
follows directly from  \cite{Bolte-1998, Bolte-1999a} and differs from the standard result only by the factor $2S+1$. We thus have
\be\label{rhoav2}
\bar\rho_\alpha(E) \sim \frac{s_\alpha}{|G|}\bar\rho(E) .
\ee
It is instructive to check that $\bar\rho(E)$ is obtained after summing over
$\rho_\alpha(E)$ with the factor $s_\alpha$ accounting for the symmetry-induced degeneracies.
Using the identity $\sum_{\alpha} s_{\alpha}^2 = |G|$, we obtain
\be
\sum_{\alpha} s_{\alpha} \bar{\rho}_{\alpha}(E) = \sum_{\alpha} \frac{s_{\alpha}^2}{|G|} \bar{\rho}(E) = \bar{\rho}(E). 
\ee

\subsection{Double groups}

\label{double}

We now turn to the case that our symmetry group is a double group. It then contains an element $\bar e$ that has no geometric component, and is typically linked to combined spin and geometrical rotation by $2\pi$. We have seen that for double groups application of the projection operator leads to zero for
the standard irreps, and that the projection operator for
the extra irreps can be written as (see (\ref{projection_double}))
\be
\hat P_\alpha=
\frac{s_{\alpha}}{|\Gamma|}\sum_{g \in \Gamma }\chi_{\alpha}(g) \U^{\dagger}(g)
\ee
where $\Gamma$ is the part of $G$ containing the geometrical symmetries.
The steps up to the trace formula (\ref{Eqn: Semiclassical dos}) then carry over with summation over $g$ restricted to  $\Gamma$. The appearances of $\frac{1}{|G|}$ appearing in intermediate steps have to be replaced by $\frac{1}{|\Gamma|}$ which  is cancelled when we sum over the $|\Gamma|$ copies of the fundamental domain.

The situation where double groups are needed leads to a slightly odd issue of interpretation. To illustrate this issue and its solution, we consider the example of a system with half-integer spin and a  symmetry with respect to rotation by $\frac{2\pi}{n}$ ($n=2,3,\ldots$) about a single axis. For such a system, we can assign to each orbit an effective winding number about the rotation axis. Now let us assume that an (unfolded) orbit winds around this axis an odd number of times. Then, intuitively, we would like to assign to this orbit the group element $\bar e$, which describes rotation be $2\pi$ as well as its odd multiples. However, at this stage orbits are only assigned elements of the geometric group $\Gamma$. As the initial and final point of the aforementioned orbit coincide, we would assign it the group element $e$.

The solution to this problem is that the assignment of $e$ or $\bar e$ does not matter for the contribution to the trace formula. In our example, the symmetry factors are $\chi_\alpha(e)=1$ and $\chi_\alpha(\bar e)=-1$, noting that only the extra irreps contribute. However due to (\ref{Galpha}), the spin factor $\tr(d_p)$ is affected by    matrices $\Us(e)=\Id$ and 
$\Us(\bar e)=-\Id$, so the differences between the two choices mutually cancel.

Hence we obtain two equivalent trace formulas, both
described by (\ref{Eqn: Semiclassical dos}). In the first, the group element $g_p$ associated to each orbit is taken from the geometric group, depending on the symmetry operation relating the initial and final point of the unfolded orbit. In the second formula, $g_p$ is taken from the double group, and it is e.g. $\bar e$ for an odd number of full rotations, or in the case
of a rotation by $2\pi+\frac{2\pi}{n}$ by $\bar e$ times the group element describing rotation by $\frac{2\pi}{n}$. Crucially, only the second formula involves group elements that are fully in line with the combination of partial rotations along the orbit. Hence, we have to use this formula if we want to write $d_p$ in the product form (\ref{Eqn: Spin precession FD}), or $g_p$ as a product of group elements along the orbit. This is why we will use this version in \cite{Blatzios-2024b}.

In the Weyl term, the replacement of $|G|$ by $|\Gamma|$ is not compensated, and we obtain\be
\label{rhoavd}
\bar{\rho}_{\alpha}(E) \sim \frac{ s_{\alpha}(2S+1)|\Omega(E)|}{|\Gamma|(2 \pi
\hbar)^f} 
\ee
and
\be\label{rhoav2d}
\bar\rho_\alpha(E)=\frac{s_\alpha}{|\Gamma|}\bar\rho(E)
\ee
instead of (\ref{rhoav}) and (\ref{rhoav2}).  In spite of this change, the average level densities of the subspectra again sum to $\bar\rho(E)$. For double groups,
  the sum is taken only over the extra irreps using that
 \be
\sum_{\alpha \ {\rm extra}} s_{\alpha}^2 = \sum_{\alpha \ {\rm standard}}
s_{\alpha}^2 = \frac{|G|}{2}=|\Gamma|,
\ee
where the two sums are equal because there is a one-to-one relation between standard and extra irreps. We thus obtain\be
\sum_{\alpha \ {\rm extra}} s_{\alpha} \bar{\rho}_{\alpha}(E) = \sum_{\alpha
 \ {\rm extra}} \frac{s_{\alpha}^2}{|\Gamma|} \bar{\rho}(E) = \bar{\rho}(E)
\ee
as desired.

Formally Eqs.\ (\ref{rhoav}) and (\ref{rhoav2}) can be regarded as special cases of (\ref{rhoavd}) and (\ref{rhoav2d}), because if $G$ is not a double group, the geometric symmetry group $\Gamma$ is naturally defined as equal to $G$.

\section{Spectral determinant}

\label{sec: spectral determinant}

In the theory of spectral statistics the spectral determinant $\Delta(E)$ plays an important role \cite{Heusler-2007, Keating-2007, Muller-2009}. It is defined as (a regularised version of) $\det( E-\hat H)$ where  $\hat H$ is the Hamiltonian. A semiclassical approximation \cite{Berry-1990,Keating-1992,Berry-1992} can be obtained by using $\det=\exp\tr\ln$, expressing $\tr\ln(E-H)$ as an integral over the trace of the Green function, and arranging the exponentiated sum over orbits as a sum over sets of orbits (so-called pseudo-orbits), denoted by $A$. Depending on whether the energy is taken with a small positive or negative imaginary part ($E^+$ vs $E^-$) and whether the spectral determinant is inverted or not we then obtain, up to a proportionality factor
\begin{align}
\Delta(E^+)&\propto
{\rm e}^{-i\pi\overline{N}(E^+)}
\sum_{A} F_A(-1)^{n_A}{\rm e}^{i S_A(E^+)/\hbar}\notag\\
\Delta(E^-)&\propto
{\rm e}^{i\pi\overline{N}(E^-)}
\sum_{A} F_A^*(-1)^{n_A}{\rm e}^{-i S_A(E^-)/\hbar}\notag\\
\Delta(E^+)^{-1}&\propto
{\rm e}^{i\pi\overline{N}(E^+)}
\sum_A F_A{\rm e}^{i S_A(E^+)/\hbar}\notag\\
\Delta(E^-)^{-1}&\propto
{\rm e}^{-i\pi\overline{N}(E^-)}
\sum_A F_A^*{\rm e}^{-i S_A(E^-)/\hbar}
\end{align}
where $\bar N(E)$ is the integrated average level density, $F_A$ is a product of stability factors $\frac{e^{-i\mu_a\frac{\pi}{2}}}{\sqrt{|\det(M_a-I)|}}$ of the orbits in $A$,  $S_A$ is the cumulative action, and $n_A$ is the number of orbits included in the pseudo-orbit $A$.
Since in our formula for the Green function (\ref{Galpha}) the stability factor is multiplied with the symmetry and spin factors, these factors carry over and lead to the replacement
\be
F_A\to F_A\prod_{a\in A}\chi_\alpha(g_a)\tr(d_a) ,
\ee
as well as to the substitution of the integrated average level density by its counterparts for our subspectrum. 

For non-inverted spectral determinants and real $E$, it is often preferable to use the Riemann-Siegel lookalike formula
\be
\label{RiemannSiegel}
\Delta(E)\propto
{\rm e}^{-i\pi\overline{N}(E)}
\sum_{A\;(T_A<T_H/2)} F_A(-1)^{n_A}{\rm e}^{i S_A(E)/\hbar}+{\rm c.c.}
\ee
Here the contribution of pseudo-orbits with cumulative periods $T_A$ larger than half of the Heisenberg time $T_H=2\pi\hbar\bar\rho$ is replaced by the complex conjugated contribution of the shorter pseudo-orbits.  Eq.\ (\ref{RiemannSiegel}) was obtained by Berry and Keating in \cite{Berry-1990,Keating-1992,Berry-1992}. This result carries over with the same replacement as above, if one uses the Heisenberg time for our subspectrum.

\section{Conclusions}

 \label{sec: conclusions}

We have reviewed the theory of irreducible representations and its applications to quantum theory, including connections to spin and time-reversal operators.
We have then established semiclassical approximations for the Green function, level density and spectral determinant of subspectra of quantum chaotic systems with symmetries and spin.
These expressions involve orbits that are periodic in the fundamental domain, with weights including the character of the correponding irrep and the trace of a matrix describing spin propagation as well as the spin components of the symmetry operators.

Our results set the stage for a study of spectral statistics in \cite{Blatzios-2024b}. As the Green function also forms the basis of semiclassical approaches to quantum transport, our results also allow to investigate transport in chaotic systems with symmetries and spin. Similar modification will carry over
to properties of many-body systems where the analogue of the chaotic  classical dynamics is the dynamics given by a chaotic nonlinear Schr\"odinger equation \cite{Engl-2014,Engl-2015,Dubertrand-2016}.

We are grateful for helpful discussions with Boris Gutkin, Charlie Johnson, Jonathan Robbins and Alexander Winter.  

\appendix 


\section{More complicated double groups}

In the main part, we gave an example for double groups based on rotations about a single axis. For an example with {\bf more than one rotation axis}, consider a system that is symmetric with respect to rotation by an angle $\pi$ about the $x$-axis (denoted by $r_x$) and rotation by an angle $\pi$ about the $y$-axis (denoted by $r_y$). Then the geometric symmetry group is given by $\Gamma=\{e,r_x,r_y,r_xr_y\}$. In case of half-integer spin, rotation by an angle $2\pi$ leads to a group element $\bar e$ whose spin matrix is $-\Id$.
The full symmetry group is then the double group $G=\Gamma\cup\bar e\Gamma$ with eight elements. One can show that a representation of this group is given by the matrices of the quaternion group $Q_8 = \{ \pm \Id, \pm i\sigma_x, \pm i \sigma_y, \pm i \sigma_z \}$ where $\sigma_x$, $\sigma_y$ and $\sigma_z$ are the Pauli matrices.

Point groups can also involve {\bf improper rotations}, including reflections and the inversion $\mathcal{I}$. If the inversion is included and $\Gamma'$ is the group of rotation symmetries, one can show that $\Gamma=\Gamma'\cup{\cal I}\Gamma'$ and $G= \Gamma'\cup{\cal
I}\Gamma' \cup \,\bar e\Gamma'\cup \bar e\mathcal{I}\Gamma'$. See \cite{Lax-1974,Doublegroups}
 for a discussion of the slightly more complicated case where the inversion is not included. 
 
\section{Spin components of non-conventional time reversal operators} 
 
We want to show that for non-conventional time reversal operators with $\hat T^2=\pm \Id$ it is justified to only consider $\Us=\Id$. 
To do so, 
we express $\Uf=\Us e^{i \pi \hat S_y/\hbar}$ in terms of generalised Pauli matrices as $\Uf=a_0\Id+ia_x\sigma_x +ia_y\sigma_y+ia_z\sigma_z $. Now unitarity implies
$\Uf\Uf^\dagger=\Id$ and for $\TR^2=\pm\Id$ we need $\Uf\Uf^*$ to be proportional to $\Id$. Hence, $\Uf(\Uf^*-\Uf^\dagger)=\Uf(-2ia_y\sigma_y)=2a_y(a_y\Id-ia_z\sigma_x -ia_0\sigma_y+ia_x\sigma_z)$ must be proportional to $\Id$, which requires either $a_y=0$ or $a_0=a_x=a_z=0$. The latter case leads to $\Us=\Id$. In the former case, the $\hat S_y$ rotation is fully removed from $\TR$. However, its effect on spin is as fundamental as the complex conjugation. (In fact, which spin components are flipped by the complex conjugation and which are flipped by a spin rotation depends on the conventions chosen for the Pauli matrices.) Hence, it is appropriate to disregard this possibility.

\section*{References}

\bibliography{trace_refs}
\bibliographystyle{iopart-num-long}

\end{document}